\documentstyle[12pt,aaspp4,epsf,epsfig]{article}

\def\lsim{\lower.5ex\hbox{$\; \buildrel < \over \sim \;$}}
\def\gsim{\lower.5ex\hbox{$\; \buildrel > \over \sim \;$}}
\def\etal{{\rm et al.}~}

\def\he#1{\hbox{$^{#1}{\rm He}$}}
\def\li#1{\hbox{$^{#1}{\rm Li}$}}
\def\b1#1{\hbox{$^{1#1}{\rm B}$}}
\def\be#1{\hbox{$^{#1}{\rm Be}$}}

\def\deg{\hbox{${}^\circ$}}

\def\beq{\begin{equation}}
\def\eeq{\end{equation}}

\def\beginapjbib{\begingroup \section*{\large \bf References}
   \parskip=.5ex plus 1.0pt
   \def\bibitem{\par \noindent \hangindent\parindent
      \hangafter=1}}
\def\endapjbib{\par \endgroup}

\begin{document}

\rightline{UMN-TH-1727/98}
\rightline{TPI-MINN-98/22}
\rightline{astro-ph/9811183}
\rightline{November 1998}

\title{The Evolution of \li6 in Standard Cosmic-Ray Nucleosynthesis}

\author {Brian D. Fields}
\affil {Department of Astronomy\\ University of Illinois\\
Urbana, IL 61801, USA}

\and

\author{Keith A. Olive}
\affil{Theoretical Physics Institute\\ School of Physics and Astronomy\\
University of Minnesota\\ Minneapolis, MN 55455, USA}

\begin{abstract}

We review the Galactic chemical evolution of \li6 and compare these
results with recent observational determinations of the lithium isotopic
ratio.  In particular, we concentrate on so-called standard Galactic
cosmic-ray nucleosynthesis in which Li, Be, and B
are produced (predominantly) by the inelastic scattering of
accelerated protons and $\alpha$'s
off of CNO nuclei in the ambient interstellar medium.
If O/Fe is constant at low metallicities, then 
the \li6 vs Fe/H evolution--as well as Be and B vs Fe/H--has 
difficulty in matching the
observations.  However, recent determinations of
Population II oxygen abundances, as measured via
OH lines, indicate that O/Fe increases at lower
metallicity; {\em if} this trend is confirmed,
then the \li6 evolution in a standard model of cosmic-ray
nucleosynthesis is consistent with the data.  We also show that another
key indicator of \li6BeB origin 
is the \li6/Be ratio which also fits the available data if
O/Fe is not constant at low metallicity.  Finally we 
note that \li6 evolution in this scenario
can strongly constrain the degree to which \li6 and
\li7 are depleted in halo stars.

\end{abstract}

\keywords{cosmic-rays -- Galaxy :abundances -- 
nuclear reactions, nucleosynthesis, abundances}

\newpage

\section{Introduction}

Both of the stable isotopes of lithium, \li6 and \li7, 
have a special nucleosynthetic status; however, their
origins are quite different. The story of \li7 is perhaps better-known,  
as this nuclide dominates by far the
Li production in big bang.  Starting in the
1980's, observations of extreme Pop II stars (Spite \& Spite 1982)
revealed a constant Li abundance versus Fe, the ``Spite plateau.''
This discovery has been confirmed many times over, and demonstrates
that Li is primordial.  Furthermore, the mean value along the plateau, 
${\rm Li/H} = (1.6 \pm 0.1) \times 10^{-10}$ (Molaro, Primas
\& Bonifacio, 1995; Bonifacio \& Molaro, 1996) agrees well with
the inferred primordial values of the other light elements,
in dramatic confirmation of big bang nucleosynthesis theory (Walker
\etal 1991; Fields
\etal 1996).   In Pop I stars, the Li abundance rises
by factor of $\sim 10$ from its primordial level and
various stellar production mechanisms have been
suggested to explain this (e.g., Matteucci, d'Antona, \& Timmes 1995).  
Here, we will focus on
the Pop II behavior of both \li7 and \li6.

Unlike \li7, the less abundant
\li6 has long been recognized as
a nucleosynthetic ``orphan''--along with Be and B,
\li6 is made neither in the the big bang nor in stars.
That is, primordial nucleosynthesis
does produces some \li6, but the abundance is unobservably
small (\S \ref{sect:prod}); moreover, stellar thermonuclear processes 
destroy \li6, whose low binding energy
renders this nucleus thermodynamically unfavorable.
Thus, \li6 can only be produced in Galactic, non-equilibrium
processes.  Just such a process was identified by
Reeves, Fowler, \& Hoyle (1970), 
in the form of Galactic cosmic ray interactions.
Specifically, the propagation of cosmic rays
(mostly protons and $\alpha$'s)
through the interstellar medium (ISM) inevitably
leads to spallation reactions on CNO nuclei
(e.g., $p+{\rm O} \rightarrow \li6$)
and fusion reactions on interstellar He
($\alpha+\alpha \rightarrow \li6$).
Furthermore, Reeves, Fowler, \& Hoyle (1970) and
Meneguzzi, Audouze, \& Reeves (1971)
showed that cosmic ray interactions 
do indeed yield solar system abundances
of these nuclides over the lifetime of the Galaxy.
This mechanism, the so-called ``Standard GCR Nucleosynthesis''
(GCRN), was thus seen to be viable.
Although lingering questions remained 
(standard GCRN alone is unable to reproduce the
solar \li7/\li6BeB ratio, nor the \b11/\b10 ratio),
this process was viewed as the conventional source of
solar system \li6BeB until the late 1980's.

This simple picture of the origin of Li has met with several complications
over the past decade or so.  There is a nagging uncertainty as
to whether or not the observed \li7 abundance in the plateau stars is in
fact representative of the primordial abundance or was partially
depleted by non-standard stellar processes. 
In other words, does the Spite plateau measure the primordial
Li abundance, or should one apply an upward correction to
offset the effects of depletion?  Stellar evolution models
have predicted depletion factors which differ widely, ranging
from essentially no depletion in standard models (for stars with $T \ga
5500$ K) to a large depletion (Deliyannis \etal 1990, 
Charbonnel \etal 1992). Depletion occurs when the base of the
convection zone sinks down and is exposed to high temperatures, $\sim 2
\times 10^6$ K for \li7 and $\sim 1.65 \times 10^6$ K for \li6 (Brown \&
Schramm 1988). Indeed, in standard stellar models, the depletion of
\li7 is always accompanied by the depletion of \li6 (though the
converse is not necessarily true).  Thus any observation of \li6 has
important consequences on the question of \li7 depletion. The issue of
depletion affects not only \li6 and BBN, but also \li6BeB and their
evolution; indeed, in this paper we will make use of this connection.

Another complication has emerged to 
apparently overthrow the picture of standard GCRN.
Namely, measurements of Be and B in halo stars
have shown that Be and B both have logarithmic slopes
versus iron which are near 1.  However, in standard
GCRN, the rates of Be and B production depend on 
the CNO target abundances.  This
implies that Be and B should be ``secondary,'' 
with abundances which vary quadratically with
metallicity; i.e., the Be and B log slopes should be 2
versus Fe.  While neutrino-process nucleosynthesis (Woosley \etal
1990; Olive \etal 1994) can help explain the linearity of B vs
Fe/H, it is the slope of [Be/H] vs. [Fe/H] which has focused the
attention of modelers of cosmic-ray nucleosynthesis. For example, to
explain the observed ``primary'' slope, new, metal-enriched cosmic ray
components have been proposed as the dominant LiBeB nucleosynthesis
agents in the early Galaxy (Cass\'{e}, Lehoucq, \& Vangioni-Flam
1995; Ramaty, Kozlovsky, \& Lingenfelter 1995; Vangioni-Flam \etal
1996; Vangioni-Flam \etal 1998a; Ramaty, Kozlovsky, Lingenfelter, \& Reeves
1997).

In the last few years, key new observations have begun to
address questions concerning the Galactic evolution and
stellar depletion of the lithium isotopes, with
the first
observations of the  \li6/\li7 ratio in a few very old halo stars 
(Smith \etal 1993,  Hobbs and Thorburn 1994, 1997).
The lithium isotopic ratio has been measured several times in HD 84937,
and most recently by Cayrel \etal (1998). The weighted average of the
available measurements yields \li6/Li = 0.054
$\pm$ 0.011  at [Fe/H] $\simeq$ -2.3.  
Recently, Smith \etal (1998) have reported
one other positive detection in BD $26\deg 3578$ with \li6/Li = 0.05
$\pm$ 0.03, at about the same metallicity. For the other halo stars 
examined, only upper limits are available. 
The rarity of \li6 detection in halo stars is
well-understood (Brown \& Schramm 1988) in terms of
\li6 depletion in stellar atmospheres.  Namely, \li6 is 
rapidly burned
at at relatively low temperatures.  Thus, \li6 survives
only in halo stars with shallow convective zones and
high effective temperatures, $T \ga 6300$ K.  
The data bear out this picture:
both \li6 detections have been in hot halo stars, and
only upper limits have been set in cooler stars.

Combined with the present ISM abundance of \li6, the halo \li6
measurements can be an important tool for testing models of Galactic
cosmic-ray nucleosynthesis.  Here, we will restrict our attention to
standard models in which accelerated protons and $\alpha$'s produced
the LiBeB elements through spallation as well as $\alpha-\alpha$ fusion
in the production of the lithium isotopes. We show that as in the case of
the evolution of Be and B, the general accord between standard GCRN and
the observational data depends crucially on the behavior of the O/Fe
ratio in Pop II (Fields \& Olive 1998).  
The model is described in detail below (\S \ref{sect:prod}); 
the basic idea is that for spallation production,
the slope vs O (rather than Fe) is the better
indicator of nucleosynthesis origin.
If O/Fe is constant, this distinction is irrelevant.  However,
new O/Fe data for Pop II reveal an evolving O/Fe ratio; 
if true, then it follows that the
Be and B slopes versus Fe are not equal to the respective slopes vs O.  
As pointed out by Fields \& Olive (1998), 
{\em if} O/Fe does vary as suggested by
Israelian et al (1998) and by Boesgaard \etal (1998), then GCRN-produced
BeB  can have a roughly ``primary'' slope
versus Fe yet a secondary slope vs O.
Within the observational errors, 
the LiBeB data vs O and Fe are consistent with 
a ``neoclassical'' or ``revised standard'' cosmic ray nucleosynthesis
consisting of (1) GCRN which makes LiBeB, without additional
metal-enriched cosmic rays, and (2) the $\nu$-process
in SNe, required to fit the solar \b11/\b10 ratio as well as the
different B and Be slopes in Pop II.
As we show below, the effect of a varying O/Fe ratio, will also
soften the slope of \li6/H vs. Fe/H, though to a lesser extent.

It is important to note that while the revised standard GCRN 
is allowed by the present data, the uncertainties in the observations
are large enough that primary models for BeB are also allowed.
Fortunately, this ambiguity will not persist:  the two classes of models
give very different predictions for LiBeB and O/Fe evolution,
and can thus be tested with more and better observations.
As described in detail below, \li6 evolution, and the
\li6/Be ratio provide one of the best discriminators between 
the different scenarios.  Below we will show in detail
the prediction of the secondary (standard) model regarding \li6, the
current data, and suggest future observations. The evolution of \li6 in a
primary model of cosmic-ray nucleosynthesis is discussed in Vangioni-Flam
\etal (1998b).

We also discuss the possibility of using this model
to constrain the degree to which both Li isotopes
are depleted in halo stars (\S \ref{sect:ev}).  
Following Steigman et al.\ (1993) and Lemoine et al.\ (1995),
we compute the maximum difference between the initial predicted \li6
and the observed \li6, and find that little \li6 depletion
is allowed. That implies that the \li7 depletion is at least
as small and probably negligible, so that the observed Spite
plateau value of Li/H should reflect the true primordial Li
abundance.

\section{The \li6 Data}
\label{sect:data}

The determination of the \li6 abundance in halo dwarf stars is
extremely difficult and requires high resolution and high signal to
noise spectra due to the tiny hyperfine splitting between the two
lithium isotopes of $\sim 0.16$ \AA. The observational challenge 
arises in part because the 
line splitting is not seen as a distinct doublet, 
but rather the narrowly shifted lines are thermally broadened so that
one sees only a single, anomalously wide absorption feature;
fits to the width of this feature are sensitive to the \li7/\li6 ratio.
The first indication of a positive
result was reported by Andersen, Gustafsson, and Lambert (1984) in the
star HD211998 with \li6/\li7 = 0.07  (though they caution
that the detection was not certain) consistent with the upper limit of
$0.1$ reported by Maurice, Spite, \& Spite (1984).  Given the relatively
low surface temperature of this star,  $T \simeq 5300$ K, standard
stellar models (e.g. Deliyannis, Demarque, \& Kawaler, 1990) would
predict that this star is severely depleted in \li6.  Indeed, the star is
depleted in \li7 ([\li7] $\simeq$ 1.22), thus lying well below the Spite
plateau.

\begin{table}[ht]
\centerline {\sc{\underline{Table 1:} Hot Halo Dwarf Stars}}

\vspace {0.1in}
\begin{center}
\begin{tabular}{|lcccc|}                     \hline \hline
Star &  Temperature (K) & [Fe/H] &  [Li]   \\ \hline
G 64-12 & $6356 \pm 75$ & -3.03 & $2.32 \pm 0.07$ & B \\
  & $6468 \pm 87$ & -3.35 & $2.40 \pm 0.07$ & IRFM \\ 
G 64-37 & $6364 \pm 75$ & -2.6 & $2.03 \pm 0.07$ & B \\
  & $6432 \pm 70$ & -2.51 & $2.11 \pm 0.06$ & IRFM \\ 
LP 608 62 & $6435 \pm 52$ & -2.51 & $2.28 \pm 0.06$ & B  \\
  & $6313 \pm 80$ & -2.81 & $2.21 \pm 0.08$ & IRFM  \\
BD 9\deg 2190 & $6452 \pm 60$ & -2.05 & $2.20 \pm 0.11$ & B  \\
  & $6333 \pm 89$ & -2.89 & $2.15 \pm 0.07$ & IRFM \\ 
BD 72\deg 94 & $6347 \pm 88$ & -1.3 &  $-$ & B \\
BD 36\deg 2165 & $6349 \pm 84$ & -1.15 &  $-$ & B \\
HD 83769 & $6678 \pm 97$ & -2.66 & $-$ & IRFM  \\
HD 84937 & $6330 \pm 83$ & -2.49 & $2.27 \pm 0.07$ & IRFM  \\
BD 26\deg 3578& $6310 \pm 81$ & -2.58 & $2.24 \pm 0.06$ & IRFM  \\
G 4-37 & $6337 \pm 92$ & -3.31 & $2.16 \pm 0.08$ & IRFM  \\
G 9-16 & $6776 \pm 84$ & -1.31 & $-$ & IRFM  \\
G 37-37 & $6304 \pm 112$ & -2.98 & $ - $ & IRFM  \\
G 201-5 & $6328 \pm 111$ & -2.64 & $2.27 \pm 0.09$ & IRFM  \\
BD 20\deg 3603 & $6441 \pm 76$ & -2.05 & $2.41 \pm 0.07$ & IRFM  \\
\hline
\end{tabular}
\end{center}
\end{table}

\indent \li6 can only be realistically expected to be
observed in stars with both high surface temperatures and at
intermediate metallicities of [Fe/H] between about -2.5 and -1.3.
Brown and Schramm (1988) determined that only in stars with surface
temperatures greater than about 6300 K will \li6 survive in the
observable surface layers of the star.  At metallicities much lower than
[Fe/H] $\sim -2.5$, the
\li6 abundance is expected to be lower due to  the short timescales
available for GCRN production. At metallicities [Fe/H] $\ga -1.3$, even
higher effective temperatures would be required to preserve \li6. 
In Table 1, we show the stellar parameters and \li7 abundances of a set of
stars from  with $T > 6300$ K from Molaro, Primas
\& Bonifacio (1995) who use the Balmer line method of Fuhrmann, Axer \&
Gehren (1994) for determining the surface temperatures (labeled B) and
from Bonifacio \& Molaro (1996) who use the Infrared Flux Method (IRFM) of
Blackwell \etal (1990) and Alonso \etal (1996).  Included in this set of
stars is the well observed HD 84937. In an early report, Pilachowski,
Hobbs, \& De Young (1989) were able to determine an upper limit of
\li6/\li7 $< 0.1$ for this hot halo dwarf star. The \li7 abundance for BD
72\deg 94 was determined by Rebolo, Molaro, \& Beckman (1988) and by 
Pilachowski, Sneden, \& Booth (1993) using other parameter choices to be
[Li] = 2.22 $\pm$ 0.09.

In a now seminal paper, Smith, Lambert, and Nissen (1993) reported the
first detection of a small amount of \li6 in the halo dwarf HD 84937 at
the level of $R \equiv \li6/(\li6+\li7) = 0.05 \pm 0.02$.  
In a slightly cooler star,
HD 19445, they found only an upper limit of $R < 0.02$.  This  observation
was confirmed by Hobbs \& Thorburn (1994, hereafter HT94) who found
\li6/\li7 $ = 0.07
\pm 0.03$ for HD 84937. HT94 observed 5 additional stars, only one of
which is in Table 1, HD 338529 also known as BD $26\deg 3578$, in which
they found the upper limit $R < 0.1$. In their sample, they list BD
$3\deg 740$ as a very hot star at 6400 K, however both the B and
IRFM methods yield lower temperatures (6264 K and 6110 K respectively). 
In contrast, HT94 did report a positive detection in HD 201891 with
\li6/\li7 $ = 0.05 \pm 0.02$, in this relatively cool (5900 K) star. In
Hobbs \& Thorburn (1997, hereafter HT97), a reanalysis of HD 84937 yielded
\li6/\li7 $ = 0.08 \pm 0.04$ and converted the detection of \li6 in HD
201891 to an upper limit $R < 0.05$. In addition, HT97 found upper limits
in 5 other cooler halo stars. 

In more recent work, Smith \etal (1998) observe nine halo stars, three of
which appear in Table 1.  In two cases, HD 84937, and BD 26\deg 3578,
they claim a positive detection for \li6 with $R = 0.06 \pm 0.03$ and $R
= 0.05 \pm 0.03$ for the two stars respectively.  In the third star, BD
20\deg 3603, \li6 was not detected and an upper limit $R < 0.02$ was
established. In the remaining six stars, no \li6 was detected with
certainty.  In addition, it is interesting to note that Smith
\etal (1998) confirm the small scatter in the \li7 abundances seen by
Molaro \etal (1995), Spite \etal (1996), and Bonifacio \& Molaro (1997),
far below that expected if \li7 were depleted in these halo stars. 

Recently, Cayrel \etal (1998) have performed new very high signal-to-noise
measurements of HD 84937 and 2 other stars, BD 36\deg 2165 and BD
42\deg 2667, the former of which is hot enough to appear in Table 1.  They
report a very accurate determination of \li6/\li7 = 0.052 $\pm$ 0.018 in
HD84937. The weighted mean of all measurements of the lithium isotopic
ratio in HD 84937 is \li6/\li7 = $ 0.057 \pm 0.012$, corresponding to $R
= 0.054 \pm 0.011$. Cayrel \etal (1998) also report a possible detection
of \li6 in the cooler of the other two stars but a lower S/N for this
observation precludes a definite detection claim. BD 42\deg 2667 was
also observed by Smith \etal (1998) with no detection reported.  In the
third star, Cayrel \etal (1998) state that no
\li6 was detected.

\section{ The Production of \li6}
\label{sect:prod}

While primordial \li7 is produced at observable levels,
\li6 production is orders of magnitude smaller
(Thomas \etal 1993; Delbourgo-Salvador \& Vangioni-Flam 1994).
For values of 
the baryon-to-photon ratio
$\eta$ consistent with \he4 and \li7, $\eta = (1.8-5) \times
10^{-10}$ (see e.g. Fields \etal 1996), 
\li6 lies in the range $\li6/{\rm H} \simeq (1.5 - 7.5) \times 10^{-14}$,
far below the levels measured in halo stars ($\sim 8 \times 10^{-12}$).
One thus infers that Li in halo stars should be dominated by the primordial
component of \li7, and the data is completely consistent with this
expectation. Furthermore, it follows that the \li6 observed in Pop II is 
due to Galactic processes.  The BBN production of \li6 was recently
reinvestigated in Nollett \etal (1997) comparing several measurements of
the important D ($\alpha, \gamma$) \li6 reaction.  By and  large they
found similar results to those in Thomas \etal (1993), showing that
the the uncertainty in this reaction could allow for perhaps a
factor 2 more \li6. 

As noted in the introduction, there have been several models
advanced to explain Pop II \li6 as well as Be and B;
all of these involve spallation/fusion of accelerated particles.  
In this paper 
we focus on the case for standard GCRN as the 
source of Pop II \li6,
in the picture of Fields \& Olive (1998),
which includes both GCRN and the $\nu$-process.
In contrast to ``primary'' models (at least in their simplest forms),
different \li6BeB nuclides  
have strongly different evolution 
in standard GCRN.
Namely, in Pop II the Li isotopes 
and \b11 ($\approx$ B) are mostly primary, due to $\alpha+\alpha$
interactions and $\nu$-process, respectively.
On the other hand, \be9 and \b10 are secondary (versus oxygen).
Consequently, the ratios of primary to secondary nuclides---e.g.,
B/Be, \li6/Be, and $\b11/\b10$---all vary strongly 
with metallicity and are good tests if measured accurately.
We note however, that the Li/Be ratio is highly model dependent as shown
in Fields, Olive, \& Schramm (1994).

The GCRN model is described in detail elsewhere
(Fields \& Olive 1998); here,
we only summarize essential cosmic ray inputs:
(1) Galactic cosmic rays are assumed to be accelerated with
the (time-varying) composition of the ambient ISM.
Consequently, interactions between cosmic ray $p,\alpha$ particles
and interstellar HeCNO dominate the LiBeB production.
(2) The injection energy spectrum is that measured for the
present cosmic rays $\propto (E+m_p)^{-2.7}$; thus the
cosmic rays are essentially relativistic, with no
large flux increase at low energies.
(3) The cosmic ray flux strength is scaled to the SN rate.
On the basis of this model, production rates are computed.
The (time-integrated) LiBeB outputs are normalized to the solar 
\li6, Be, and \b10 abundances, the isotopes that 
are of exclusively GCR origin.  We adopt the solar abundances 
and isotopic ratios of Anders \& Grevesse (1989) for all but elemental
B, which is taken from the more recent
determination of Zhai \& Shaw (1994). 
The scaling factor effectively measures the average Galactic flux today,
and is calculated from an unweighted average of the 
each of the scalings for \li6, Be, and \b10.

To follow the LiBeB evolution in detail, the
production rates are incorporated into a simple
Galactic chemical evolution code.
Closed and open box models are both able to
give good results; here, we will adopt a closed box
for simplicity.
The model has an initial mass function 
$\propto m^{-2.65}$, and uses 
supernova yields due to Woosley \& Weaver (1995).
In this model, O/Fe indeed varies in Pop II. 
However, with the Woosley \& Weaver (1995) yields,
the model cannot reproduce a slope for O/Fe vs.\ Fe/H
as steep as that observed by Israelian \etal (1998) and
Boesgaard \etal (1998).\footnote {
Note that Chiappini, Matteucci, Beers, \& Nomoto (1998)
also find a changing O/Fe in Pop II .  Their results are
roughly consistent with ours when using the Woosley \& Weaver (1995)
yields, and they indeed find a 
a steeper [O/Fe] variation when using 
the Thielemann, Nomoto, \& Hashimoto (1996) yields. 
}
Since the O-Fe behavior is crucial, and the Fe yields
are the more uncertain (due to, e.g.,
the dependence on the Type II supernova mass cut
as well as the inclusion of Type Ia supernova yields) 
we use the O trends as calculated in the model, 
but scale Fe from the O outputs and the observed O/Fe logarithmic slope.
For comparison, we will present results from
a model with the na\"{\i}ve scaling Fe $\propto$ O,
to show the effect of variations in the Pop II slope of O/Fe.
In both cases, we use the observed Pop I relation
$[{\rm O}/{\rm Fe}] \simeq -0.5 [{\rm Fe}/{\rm H}]$ over the range
$[{\rm Fe}/{\rm H}] > -1$.

\section {Galactic evolution of \li6}
\label{sect:ev}

Before presenting model results, some discussion is in order
regarding how to compare the theory with the data.
Since stellar \li6 abundances may have suffered
some depletion, the observed \li6 represents a firm
a lower limit on the initial abundance. 
For an evolution model to be acceptable, 
it must therefore predict a \li6 abundance which
lies at or above the observed levels (within errors).
If a model is viable by these criteria, then
the difference between the theory and the data
quantifies the possible depletion.

\begin{figure}[htb]
\epsfysize=7.5truein
\epsfbox{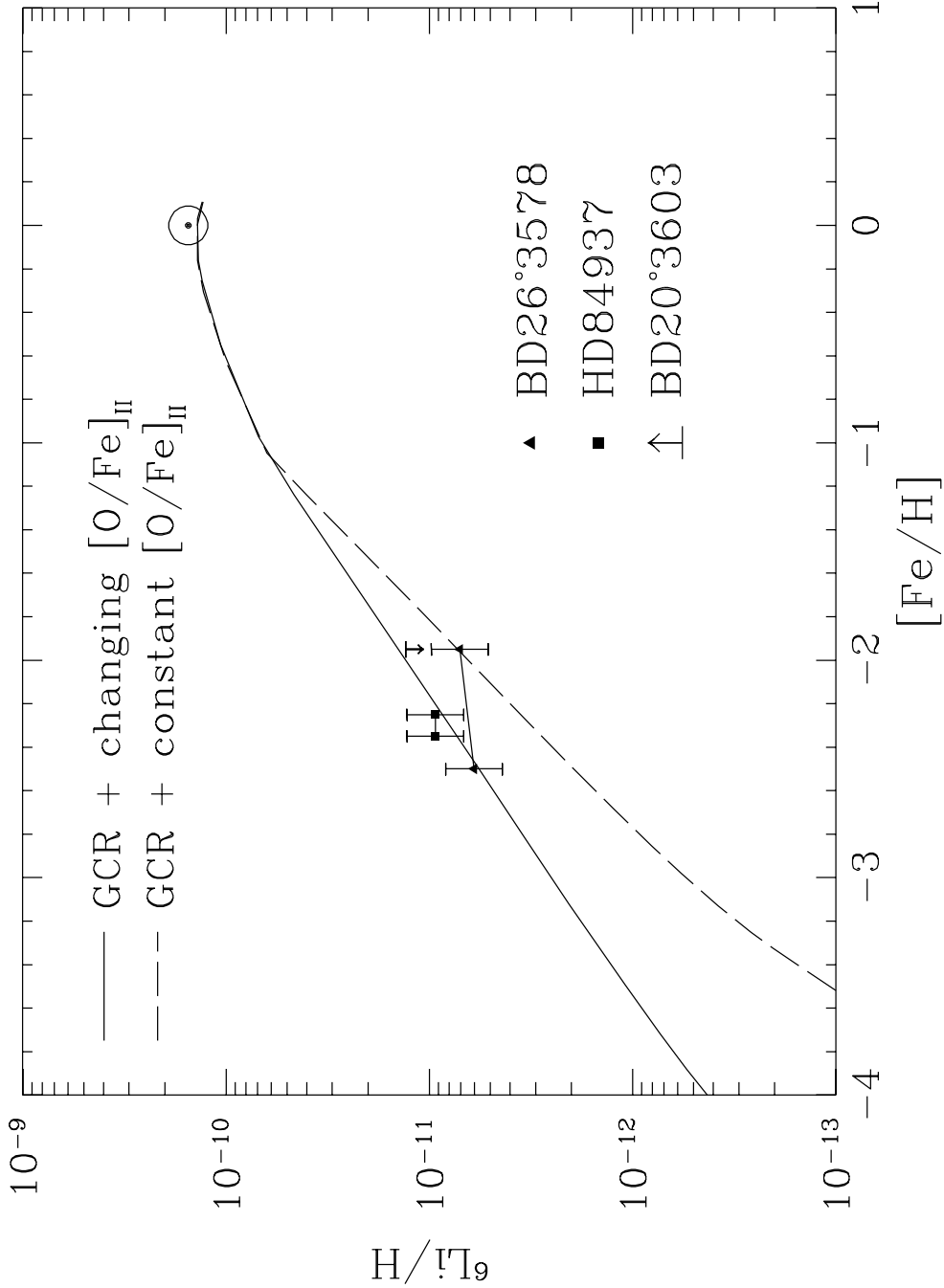}
\caption{The \li6 evolution as a function of [Fe/H].
{\it Solid line}:  the ``revised standard'' GCRN model.
Here Fe is
scaled from the calculated O to fit the observed [O/Fe]--[Fe/H] slope.
{\it Dashed line}:  the GCRN model with 
Fe $\propto$ O in POP II. The error bars on the points are 2 sigma
errors, and the spread in the points connected by lines show the
uncertainty due to stellar parameter choices.}
\label{fig:li6-fe}
\end{figure}

Model results for \li6 vs Fe appear in 
Figure \ref{fig:li6-fe}, for
[O/Fe]-[Fe/H] Pop II slope 
$\omega_{\rm O/Fe} = -0.31$ (the proposed GCRN model)
and $\omega_{\rm O/Fe} = 0$ for comparison.
We see that GCRN does quite
well in reproducing both solar and Pop II \li6
when O/Fe is allowed to evolve in Pop II.
On the other hand, if O/Fe is constant, 
then the \li6-Fe slope is steeper and 
the model underproduces the Pop II \li6.
Clearly, the O/Fe behavior in Pop II is crucial to determine accurately.
Note that because of the large uncertainty in [Fe/H] for  BD 26\deg 3578,
this star does not at this time provide a stringent constraint. 

\begin{figure}[htb]
\epsfysize=7.5truein
\epsfbox{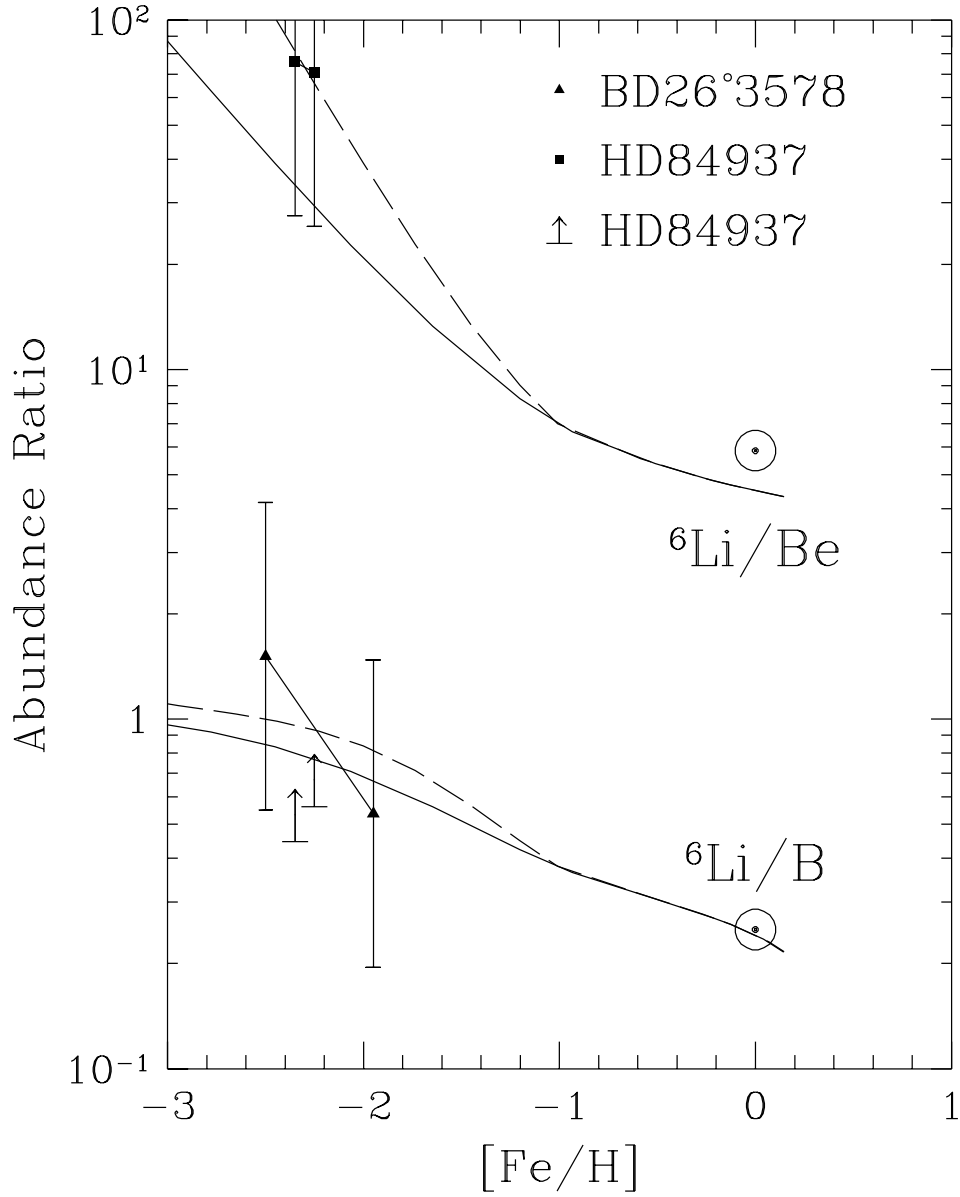}
\caption{The evolution of the \li6/Be and \li6/B ratios.  
Models are as in Figure \protect\ref{fig:li6-fe}. The error bars on the
points are 2 sigma errors, and the spread in the points connected by
lines show the uncertainty due to stellar parameter choices.}
\label{fig:li6_beb-fe}
\end{figure}

Another test of the GCRN
model is to compare primary versus
secondary nuclides, e.g., \li6/Be.  
Figure \ref{fig:li6_beb-fe} plots \li6/Be and \li6/B vs Fe
for the two O/Fe models.
We see that while the model with changing O/Fe is consistent with
\li6/Be for solar and Pop II metallicities,
the uncertainty in the data do not sufficiently discriminate between
this and the model with constant Pop II O/Fe. However, in purely primary
models (with constant O/Fe), one expects \li6/Be to be approximately
constant with
respect to [Fe/H] in Pop II, and that is clearly disfavored by the data,
albeit there is only one star with both \li6 and Be determined. 
 While there is no positive
detection of B/\li6 in a halo star, the figure makes clear that this
ratio is also a good test of the model.  In Pop II, both ${\rm B} \approx
\b11$ and \li6 are dominated by primary processes 
($\nu$-process and $\alpha+\alpha$, respectively),
and thus the \li6/B ratio changes much less strongly
than does \li6/Be. 
Note that in our scenario, \li6 and Be are pure cosmic ray
products.  Consequently, the \li6/Be curve is 
a particularly clean prediction of the model,
free of any normalization between different sources;
indeed, even the normalization of the present-day cosmic ray
flux strength drops out as a common factor.
By contrast, the  B/\li6 ratio does depend on the relative normalization
between the GCRN and $\nu$-process yields (which is fixed so that
\b11/\b10 equals the solar ratio 4.05 at [Fe/H] = 0).

The results shown above indicate that the standard cosmic-ray origin for
\li6BeB is in fact consistent with the data.  This is contrary to the
conclusions of Smith, Lambert \& Nissen (1998), who concluded that on the
basis of the solar ratio of
$\li6/{\rm Be}_\odot = 5.9$ and the value of this
ratio for HD84937, $\li6/{\rm Be} \simeq 80 \gg \li6/{\rm Be}_\odot$,
an additional source of \li6 was necessary. 
This conclusion assumes the observed linear evolution of [Be/H] vs.
[Fe/H] and the expected linear evolution of \li6 as a primary element due
to $\alpha-\alpha$ fusion.  In this case one would expect the \li6/Be
ratio to be constant, which from a simple examination of
Figure \ref{fig:li6_beb-fe} is
clearly not the case.  In standard GCRN (with constant O/Fe at low
metallicities), \be9 is a secondary isotope, and given the linearity of
[\li6], one should expect that
\li6/Be is inversely proportional to Fe/H (i.e., to have
a log slope of -1).  However, if we take [O/Fe]
= $\omega_{\rm O/Fe}$[Fe/H], then we would expect up to an additive
constant (Fields \& Olive 1998) 
\beq
[{\rm Be}]  = 2 (1 + \omega_{\rm O/Fe}) \ {\rm [Fe/H]}
\eeq 
and
\beq
[\li6]  =  (1 + \omega_{\rm O/Fe}) \ {\rm [Fe/H]}
\eeq
so that
\beq
[\li{6}/{\rm Be}] = - (1 + \omega_{\rm O/Fe}) \ {\rm [Fe/H]}
\eeq
Now for the Israelian et al.\ (1998) value of 
$\omega_{\rm O/Fe} = -0.31$, we would predict a dependence which
is consistent with the data as shown in Figure \ref{fig:li6_beb-fe}.
(The Boesgaard \etal (1998) value is very similar, $\omega_{\rm O/Fe} =
-0.35$.) While the case for a nonzero O/Fe Pop II slope has not been
conclusively made, it is nevertheless striking that
the reported $\omega_{\rm O/Fe}$ can explain all of the observed
\li6, Be, and B evolution within a simple (and canonical!) model.

The data as shown in the figures and compared to the models do not
take into account any depletion of \li6.  To be sure, there is still a
great deal of uncertainty in the amount of depletion for both \li6 and
\li7 as well as the relative depletion factor, $D_6/D_7$
(Chaboyer 1994, Vauclair and Charbonnel 
1995, Deliyannis \etal 1996, Chaboyer 1998, Pinsonneault \etal
1992, Pinsonneault \etal 1998). In the remainder of this paper we examine
to what extent, the data (present or future) can tell us about the degree
to which the lithium isotopes have been depleted and hence the
implications for the primordial abundance of \li7.  In addition, we will
show that future data on the lithium isotopic ratio may go a long way in
resolving some of the key uncertainties in GCRN.

The observed lithium abundance can be expressed as 
\beq
{\rm Li}_{\rm Obs} = D_7 ( {\rm \li7_{BB}} + {\rm \li7_{CR}}) + 
D_6 ( {\rm \li6_{BB}} + {\rm \li6_{CR}})
\label{li}
\eeq
where the $D_{6,7} < 1$ are the \li{6,7} depletion factors.
Ignoring the depletion factors for the moment, we see that lithium (and
in particular \li7) has two components, due to
big bang and cosmic ray production. In principle, given a model
of cosmic-ray nucleosynthesis, one could use the observed Be abundances in
halo stars along with the model predictions of Be/Li and \li6/\li7 to
extract a cosmic-ray contribution to \li7 and through (\ref{li})
the big bang abundance of \li7 (Walker \etal 1993, Olive \& Schramm 1992).
Unfortunately this procedure is very model-dependent
since Li/Be can vary between 10 and $\sim 300$
depending on the details of the cosmic-ray sources and
propagation--e.g., source spectra shapes, escape pathlength
magnitude and energy dependence, and kinematics
(Fields, Olive \& Schramm 1994)
On the other hand, the \li6/\li7 ratio is a relatively 
model-independent prediction of cosmic-ray nucleosynthesis. 
With more data, one could use this model-independence to great advantage,
as follows.
Given enough \li6 Pop II data, one could use the observed \li6 evolution
(1) to infer $\li7_{\rm CR}$ and thus $\li7_{\rm BB}$,
and (2) to measure \li6/Be and thereby constrain in more detail the 
nature of early Galactic cosmic rays.

In standard stellar models, Brown \& Schramm (1988) have argued that $D_6
\sim D_7^\beta$ with $\beta \approx 60$.  Clearly in this case any
observable depletion of \li7 would amount to the total depletion of \li6. 
Hence the observation of \li7 in HD84937 has served as a basis to limit
the total amount of \li7 depletion (Steigman \etal 1993, Lemoine \etal
1997, Pinsonneault \etal 1998). There are however, many models based
on diffusion and/or rotation which call for the depletion \li6 and \li7
even in hot stars. 
The weakest constraint comes from assuming that depletion
occurs entirely due to mixing, so the destruction of
the Li isotopes is the same despite the greater fragility of 
\li6.
Because \li6/\li7 $\sim 1$ in cosmic-ray
nucleosynthesis, the observation of \li6 does exclude any model with
extremely large \li6 depletion on the basis of the Spite plateau for \li7
up to [Fe/H] = -1.3 (Pinsonneault \etal 1998, Smith \etal 1998).
However, barring an alternative source for the production of \li6, the
data are in fact much more restrictive. At the 2$\sigma$ level, the model
used to produce the evolutionary curve in Figure \ref{fig:li6-fe}, 
would only allow a
depletion of \li6 by 0.15 dex ($D_6 > 0.7$); since $D_7 \ge D_6$, this
is also a lower limit to $D_7$.  
We note that with improved data on BeB as well, and knowing that $D_{\rm
B} \ge D_{\rm Be} \ge D_7 \ge D_6$, one can further limit the degree of
depletion in the lighter isotopes.

Further constraints on $D_7$ become available if we adopt a
model which relates \li6 and \li7 depletion.  E.g., 
if we use $\log D_6 = -0.19 + 1.94 \log
D_7$ as discussed in Pinsonneault \etal (1998), the data in the context
of the given model would not allow for any depletion of \li7.  Of course
there is uncertainty in the model as well.  Using the Balmer line stellar
parameters, we found  (Fields \& Olive 1998) $\omega_{\rm O/Fe} = -0.46
\pm 0.15$.  Using the value of -0.46, we determine that at the 2$\sigma$
(with respect to the \li6 data) that $\log D_6 > -0.32$ and would still
limit $\log D_7 > -0.07$. Even under what most would assume is an extreme
O/Fe dependence of $\omega_{\rm O/Fe} = -0.61$,  \li6 depletion is
limited to by a factor of 3.5 and corresponds to an upper limit on the
depletion of \li7 by 0.2 dex.  This is compatible with the upper limit in
Lemioine \etal (1997) though the argument is substantially different.

It should be clear at this point, that improved ($\equiv$ more) data on
\li6 in halo stars can have a dramatic impact on our understanding of
cosmic-ray nucleosynthesis and the primordial abundance of \li7.
Coupled with improved data on the O/Fe ratio in these stars, we would be
able to critically examine these models on the basis of their predictions
of \li6 and \be9.

\section{Conclusion}
\label{sect:conclude}

We have considered the evolution of 
\li6 in context of standard Galactic cosmic ray nucleosynthesis.
In this scenario, \li6 and \li7 have a primary origin,
due to the dominance of $\alpha+\alpha$ in Pop II,
while Be and \b10 are secondary (with \b11 primary due to
the neutrino process).
\li6 thus provides an excellent 
diagnostic of LiBeB origin, both by itself and in
ratio to Be and B. 
We find that if O/Fe has a changing slope in Pop II,
as suggested by Israelian et al.\ (1998) and Boesgaard \etal (1998), then
standard GCRN provides a good fit to \li6/H and \li6/Be
for both Pop II and solar data.
On the other hand, a model with constant O/Fe in Pop II
does poorly, illustrating the need to determine
the O/Fe trend accurately.

Given the evolution scheme proposed here, 
one can constrain both \li6 and \li7 depletion
in halo stars.  The predictions here are 
in good agreement with the observed \li6 data, 
uncorrected for depletion; it follows that in our
model, the \li6 depletion cannot be very large:
the abundance is reduced by a factor of 3.5 at the extreme,
and more likely a factor of $<2$.  Using the model
discussed in Pinsonneault \etal (1998), this
leads to an upper limit on \li7 depletion of 0.2 dex.

It is interesting to note the robustness of our conclusions
regarding \li6 evolutionary constraints Pop II Li depletion.
As noted above, the GCRN model is not the only possible scenario
for LiBeB production allowed by the current data. 
A class of sharply different scenarios is also viable,
in which all of LiBeB are primary products through new
mechanisms in addition to the standard GCRN.  
The \li6 evolution in one such model is considered 
in detail by Vangioni-Flam et al.\ (1998b), in an
analysis very similar to our own.  
Interestingly, the two very different models get similarly
strong constraints.
Thus the basic conclusion is quite robust that viable
LiBeB evolution models
imply small \li6 depletion. 

We wish to re-emphasize the utility of, and need for, more 
and better observations
of \li6, Be, and B in Pop II.  
The ambiguity of the putative ``primary'' versus ``standard GCRN''
scenarios can be resolved with careful observations, which
will also pave the way for sharper tests of Li depletion,
a better knowledge of the primordial Li abundance,
and a better understanding of early Galactic cosmic rays.

Finally, as this volume celebrates the life and science of 
David Schramm, it is particularly fitting to
point out his major role in the study of LiBeB origin and evolution.  
A single example of his impact is the prescient work
of Reeves, Audouze, Fowler, \& Schramm (1973), 
which sweepingly laid out a paradigm for the origin of
the light elements.  
The LiBeB origin proposed by Reeves et al.\ 
combined contributions from primordial \li7,
cosmic-ray-produced \li6, Be, and \b10, 
and an additional stellar \li7 and \b11 source.
This basic picture has served as the
starting point for all subsequent work in the field,
including the model presented here.

\acknowledgements 
We are grateful to Elisabeth Vangioni-Flam and Michel Cass\'{e}
for many useful discussions and comments on an earlier 
version of this work. 
We would also like to dedicate this work in the memory of David Schramm,
an advisor to us both.  He brought to
the area of cosmic-ray nucleosynthesis the same
insight and unparalleled enthusiasm he showed
for all of his many research interests. 
This work was supported in part by
DoE grant DE-FG02-94ER-40823 at the University of Minnesota.

\beginapjbib

\bibitem Anders, E. and Grevesse, N. , 1989, Geochim. Cosmochem. Acta,
 53, 197 
\bibitem Andersen, J., Gustafsson, B., \& Lambert, D.L. 1984, A \& A,
136,75
\bibitem Alonso, A., Arribas, S., \& Martinez-Roger, C. 1996, A \& AS,
117, 227

\bibitem Blackwell, D.E., \etal 1990, A \& A, 232,396

\bibitem Boesgaard, A.M., King, J.R., Deliyannis, C.P., \& Vogt, S.S.
1998, AJ, submitted

\bibitem Bonifacio, P. \& Molaro, P. 1997, MNRAS, 285, 847
\bibitem Brown, L. \& Schramm, D.N. 1988, ApJ, 329, L103

\bibitem Cass\'{e}, Lehoucq, R., \& Vangioni-Flam, E.
1995, Nature, 373, 318

\bibitem Cayrel, R., Spite M., Spite F., Vangioni-Flam, E., Cass\'e,
 M. and Audouze, J. 1998, AA submitted

\bibitem Chaboyer, B. 1994, ApJ, 432, L47
\bibitem Chaboyer, B. 1998, submitted,  astroph/9803106
\bibitem Charbonnel, C., Vauclair, S. \& Zahn, J.P. 1992, AA, 255, 191
\bibitem Chiappini, C., Matteucci, F., Beers, T.C., \& Nomoto, K. 1998, 
  ApJ, in press (astro-ph/9810422)
\bibitem Delbourgo-Salvador, P. \& Vangioni-Flam, E. 1994, in " Origin and
 Evolution of Elements", Edts Prantzos \etal, Cambridge University Press, p. 52
\bibitem Deliyannis, C.P., Demarque, C. \& Kawaler, S.D. 1990, ApJS 73, 21

\bibitem Deliyannis, C.P., King, J.R. and Boesgaard, A.M., 1996, 
BAAS, 28, 916

\bibitem Fields, B.D., Kainulainen, K., Olive, K.A. \& Thomas, D. 1996, 
New Astronomy, 1, 77

\bibitem Fields, B.D., \& Olive, K.A. 1998, ApJ, submitted
(astro-ph/9809277)

\bibitem Fields, B.D., Olive, K.A., \& Schramm, D.N.
1994, ApJ, 435, 185

\bibitem Fuhrmann, K., Axer, M., \& Gehren, T. 1994, A \& A, 285, 585

\bibitem Israelian, G., Garc\'{\i}a-L\'{o}pez, \& Rebolo, R.
1998, ApJ, 507, 805

\bibitem Hobbs, L.M. \& Thorburn, J.A. 1994, ApJ, 428, L25 (HT94)

\bibitem Hobbs, L.M. \& Thorburn, J.A. 1997, ApJ, 491, 772 (HT97)

\bibitem Lemoine, M., Schramm, D.N., Truran, J.W., \& Copi, C.J. 
1997, ApJ, 478, 554

\bibitem Matteucci, F., d'Antona, F. \& Timmes, F.X. 1995, A\&A, 303, 460

\bibitem Maurice, E., Spite, F., \& Spite, M. 1984, A \& A, 132, 278 

\bibitem Meneguzzi, M., Audouze, J. \& Reeves, H. 
1971, A\&A, 15, 337

\bibitem Molaro, P., Primas, F., \&  Bonifacio, P. 1995, A \& A,
295, L47 

\bibitem Nollett, K.M., Lemoine, M. \& Schramm, D.N. 1997, Phys. Rev.
C56, 1144

\bibitem Olive, K.A., Prantzos, N., Scully, S., \&
Vangioni-Flam, E. 1994, ApJ, 424, 666

\bibitem Olive, K.A. \& Schramm, D.N. 1992, Nature, 360, 439

\bibitem Pilachowski, C.A., Hobbs,, L.M., \& De Young, D.S. 1989, ApJ,
345, L39

\bibitem Pilachowski, C.A., Sneden, C., \& Booth, J. 1993, ApJ, 407, 699

\bibitem Pinsonneault, M.H., Deliyannis, C.P. \&
 Demarque, P. 1992, ApJS, 78, 181

\bibitem Pinsonneault, M.H., Walker, T.P., Steigman, G. \& Naranyanan, V.K.
1998, ApJ submitted

\bibitem Ramaty, R., Kozlovsky, B., Lingenfelter, R.E.,
1995, ApJ, 438, L21

\bibitem Ramaty, R., Kozlovsky, B., Lingenfelter, R.E., \& Reeves, H.
1997, ApJ, 488, 730

\bibitem Rebolo, R., Molaro, P., \& Beckman, J.E. 1988, A \& A, 192, 192

\bibitem Reeves, H., Audouze, J., Fowler, W.A., \& Schramm, D.N. 
   1973, ApJ, 177, 909

\bibitem Reeves, H., Fowler, W.A., \& Hoyle, F. 1970, Nature, 226, 727

\bibitem Smith, V.V., Lambert D.L. \& Nissen P.E. 1993, ApJ, 408, 262

\bibitem Smith, V.V., Lambert, D.L. \& Nissen, P.E., 1998, ApJ, 506, 405

\bibitem Spite, F. \& Spite M. 1982, AA, 115, 357

\bibitem Spite, M. Francois, P., Nissen, P.E., \& Spite, F. 1996, 
A \& A, 307, 172

\bibitem Steigman, G., Fields, B.D., Olive, K.A., Schramm, D.N., \&
Walker, T.P. 1993, ApJ, 415, L35

\bibitem Thielemann, F.-K., Nomoto, K., \& Hashimoto, M. 1996, ApJ, 460, 408

\bibitem Thomas, D., Schramm, D.N., Olive, K.A. \& Fields, B.D., 
1993, ApJ, 406, 569 

\bibitem Vangioni-Flam, E., Cass\'e, M., Cayrel, R., Audouze, J., 
  Spite, M., \& Spite, F. 1998b, New Astronomy, submitted

\bibitem Vangioni-Flam, E., Cass\'e, M., Olive, K. \& Fields B.D. 
1996, ApJ 468, 199

\bibitem Vangioni-Flam, E. Ramaty, R., Olive, K.A. \& Cass\'e, M. 1998a,
A \& A, 337, 714

\bibitem Vauclair, S. \& Charbonnel, C. 1995, A \& A, 295, 715

\bibitem Walker, T.P., Steigman, G. Schramm, D.N., Olive, K.A., \&
Fields, B.D. 1993, ApJ, 413, 562

\bibitem Walker, T.P., Steigman, G. Schramm, D.N., Olive, K.A., \&
Kang, K. 1991, ApJ, 376, 51

\bibitem Woosley, S.E., Hartmann, D., Hoffman, R., Haxton, W.
1990, ApJ., 356, 272

\bibitem Woosley, S.E. \& Weaver, T.A. 1995, ApJS,  101, 181

\bibitem Zhai, M., \& Shaw, D. 1994, Meteoritics, 29, 607

\endapjbib

\newpage

\end{document}